\documentclass[%
 prl,preprint,groupedaddress,
 amsmath,amssymb,
 aps,
]{revtex4-1}

\usepackage{graphicx}% Include figure files
\usepackage{dcolumn}% Align table columns on decimal point
\usepackage{bm}% bold math
\begin{document}

%\preprint{APS/123-QED}

\title{Differentiating the Higgs boson from the dilaton and the radion at hadron colliders}% breaks with \\
%\thanks{A footnote to the article title}%

\author{Vernon Barger, Muneyuki Ishida$^\dagger$, and Wai-Yee Keung$^\ddagger$}
% \homepage{http://www.Second.institution.edu/~Charlie.Author}
% \email{barger@wisc.edu}
\affiliation{
Department of Physics, University of Wisconsin, Madison, WI 53706, USA\\
$^\dagger$ Department of Physics, Meisei University, Hino, Tokyo 191-8506, Japan\\
$^\ddagger$ Department of Physics, University of Illinois, Chicago, IL 60607, USA
}%

%\author{Muneyuki Ishida}
%\altaffiliation[ ]{Department of Physics, University of Wisconsin-Madison. A visitor until March 2012.}%Lines 
% \email{mishida@wisc.edu}
%\affiliation{%
%Department of Physics, Meisei University, Hino, Tokyo 191-8506, Japan
%}%

%\collaboration{MUSO Collaboration}%\noaffiliation

%\author{Wai-Yee Keung}
% \homepage{http://www.Second.institution.edu/~Charlie.Author}
% \email{keung@uic.edu}
%\affiliation{
%Department of Physics, University of Illinois, Chicago, IL 60680, USA
%}%
%\affiliation{
% Third institution, the second for Charlie Author
%}%
%\author{Delta Author}
%\affiliation{%
% Authors' institution and/or address\\
% This line break forced with \textbackslash\textbackslash
%}%

%\collaboration{CLEO Collaboration}%\noaffiliation

\date{\today}% It is always \today, today,
             %  but any date may be explicitly specified

\begin{abstract}
A number of candidate theories beyond the standard model (SM) predict 
new scalar bosons below the TeV region. Among these,
the radion, which is predicted in the Randall-Sundrum model, and the dilaton,
which is predicted by the walking technicolor theory, 
have very similar couplings to those of the SM Higgs boson,
and it is very difficult to differentiate these three spin-0 particles 
in the expected signals of the Higgs boson at the LHC and Tevatron.
We demonstrate that
the observation of the ratio $\sigma(\gamma\gamma)/\sigma(WW)$ gives a simple and decisive way,
independently of the values of model parameters: 
the VEVs of the radion and dilaton fields. 
\end{abstract}

\pacs{14.80.Bn 14.80.Ec}% PACS, the Physics and Astronomy
                             % Classification Scheme.
%\keywords{Suggested keywords}%Use showkeys class option if keyword
                              %display desired
\maketitle

%\tableofcontents
A number of candidate theories beyond the Standard Model (SM) predict 
new scalar bosons below the TeV region. 
When a scalar boson signal is detected in the Higgs search at the LHC,  
it is very important to determine whether it is really a SM Higgs boson
or another exotic scalar.
Among these, the radion($R$), predicted in the Randall-Sundrum (RS) 
model\cite{RS,RS1,RS6,GRW,CHL,RS2,RS3,RS4,RS5,ADMS,ACP,CGPT,CGK,Han,Dav,Dav2,Toharia,BI}, and the dilaton($D$),
predicted in spontaneous $scale$ $symmetry$ 
breaking\cite{GGS,Fujii,WT0,WT2,WT3,WT4,Sannino},
have very similar couplings to those of the standard model Higgs boson ($H$), and it is very difficult 
to differentiate these three particles, $DHR$, in the signals. 
A distinctive difference\cite{GGS,Toharia,BIK,Logan} is in their couplings to massless gauge bosons.
We demonstrate that
the ratios $\sigma(\gamma\gamma)/\sigma(WW)$ are different from each other, and 
their observation gives a decisive method to distinguish these three spin-0 particles. 
Our main result is given in Fig.~\ref{fig2}.
It is important that the $\sigma(\gamma\gamma)/\sigma(WW)$ ratio is independent of the model-paramneters; 
the VEVs of the radion and dilaton fields.
The test applies to both LHC and Tevatron experimental searches.

For definiteness we consider the dilaton coupling given in ref.\cite{GGS}, 
which is the same as the dilaton coupling in 4-dimensional walking technicolor theory\cite{WT0,WT2,WT3,WT4,Sannino}
where all SM fields are composites of strongly interacting fields in conformal field theory (CFT). 
In AdS$/$CFT correspondence this dilaton is dual to the radion\cite{RS1,RS6,GRW} in 
the original Randall-Sundrum (RS1) model\cite{RS},
where all the SM fields are localized at the infrared (IR) brane in the 5-dimensional 
Anti-de Sitter(AdS) space background. 
We consider the radion coupling of the Randall-Sundrum (RS2) model 
given in ref.\cite{CHL} where all the SM fields are 
in the bulk\cite{RS2,RS3,RS4,RS5,ADMS,ACP,CGPT,CGK,Han,Dav,Dav2,Toharia,BI}.
% Generically,
%the radion has flavor changing neutral currents (FCNC)\cite{Toha2} and 
The radion has bulk couplings to the gauge bosons.  
It is dual\cite{CHL} to the dilaton in CFT.
%, where light quarks are elementary and heavy quarks are composites. 
We do not consider flavor changing neutral current (FCNC) processes; however,
we note that a dilaton in a particular CFT with SM fields
that are elementary and weakly coupled can generically have FCNC\cite{refe1,refe2}, 
as can the radion of the RS2 model\cite{Toha2} considered here.    
We will study collider signatures from the gauge coupling differences of $DHR$ in the following.

\noindent\underline{\it Effective Lagrangians}\ \ \ 
We treat the SM Higgs boson $H$, the radion $R$, and the dilaton $D$, which are also denoted
as $(\ \varphi^i\ )=(\varphi^1,\varphi^2,\varphi^3)=(h^0,\phi,\chi)\equiv (H,R,D)$.
The vacuum expectation values (VEV) of these fields are denoted as 
\begin{eqnarray}
(\ F_i\ ) &=& (F_1,F_2,F_3)=(-v,\Lambda_\phi,f)
\label{eq1}
\end{eqnarray}
for $H$, $R$, and $D$, respectively.
The $F_i^{-1}$ determine overall coupling strengths of these particles, and $F_1=-v=-246$~GeV. 

The Lagrangian $L_{\rm eff}$ of the interactions\cite{RS6,GRW,GGS,Han,CHL,Dav2} with the SM particles is given by 
\begin{eqnarray}
L_{\rm eff} &=& L_A + L_V + L_{f} + L_h + L_{AV}
\label{eq2}\\
L_A &=& -\frac{\varphi^i}{4 F_i}\left[
\left( \frac{1}{kL}+\frac{\alpha_s}{2\pi}b_{QCD}^i  \right)\sum_a F_{\mu\nu}^a F^{a\ \mu\nu}
+ \left( \frac{1}{kL}+\frac{\alpha}{2\pi}b_{EM}^i  \right) F_{\mu\nu} F^{\mu\nu}
 \right] 
\label{eq3}\\
L_V &=& -\frac{2 \varphi^i}{F_i}\left[
\left( m_W^2 W^+_\mu W^{-\mu}+\frac{1}{4kL}W^+_{\mu\nu}W^{-\mu\nu}  \right)
+ \left( \frac{m_Z^2}{2} Z_\mu Z^\mu +\frac{1}{8kL}Z_{\mu\nu}Z^{\mu\nu} \right) 
 \right] 
\label{eq4}\\
L_{AV} &=&  -\frac{\varphi^i}{F_i}
 \frac{\alpha}{4\pi}\ b_{Z\gamma}^i\  F_{\mu\nu} Z^{\mu\nu}
\label{eq5}\\
L_{f} &=& \frac{\varphi^i}{F_i}\sum_f I_f^i m_f \bar ff
\label{eq6}\\ 
L_h &=& \frac{\varphi^i}{F_i}(2m_h^2 h^2-\partial_\mu h \partial^\mu h)
\label{eq7}
\end{eqnarray}
where $L$ denotes the separation of the branes in the RS2 model and $kL$ is a parameter 
that governs the weak scale-Planck scale hierarchy.
The $1/kL$ term is absent for the dilaton and the Higgs boson.
$L_A$ specifies the couplings to the massless gauge bosons, and 
$F_{\mu\nu}^a(F_{\mu\nu})$ represents the field strength of gluon(photon).
$L_V$ gives the couplings to the weak bosons, and 
$W^+_{\mu\nu}=\partial_\mu W^+_\nu -\partial_\nu W^+_\mu$ etc.
In the fermion-coupling Lagrangian $L_f$, the factors are $I_f^{H}=-1$ and $I_f^D=1$ for all fermions $f$,
while the $I_f^R$ depend upon the bulk wave functions of the fermion $f$ in the RS2 model.
We take $I_b^R =1.66$ for one value of $b\bar b$ coupling\cite{Dav} as our example.
$L_h$ represents the couplings to the Higgs boson. 
$L_h$ is also applicable to the $\varphi^i=R,D$. 
In the radion effective interaction, the brane kinetic terms are taken to be zero\cite{Dav2}. 

A distinction in Eq.~(\ref{eq3}) is the $gg$ and $\gamma\gamma$ couplings 
$\frac{1}{kL}+\frac{\alpha_{s,EM}}{2\pi}b_{QCD,EM}^i$. 
Their expressions are given in Table~\ref{tab1}.
$b_{QCD,EM}^i$ is given by the sum of the triangle-loop
contributions of top quark and $W$ boson and the $\beta$ function coefficient
appearing in the trace-anomaly of the SM energy-momentum tensor $T_{\mu\nu} (SM)$\cite{RS1,GRW,Sannino,GGS}.
The trace-anomaly term contributes for $R$ and $D$ but not for $H$.
Here we should note that 
the $\beta$ function contributions (the second column) always count all favors "light or heavy". 
But, the mass-coupling-term of the triangle-loop diagram operates in a way to cancel the heavy countings 
if the $D$ (or $R$) masses are lower than the corresponding threshold.
As a result, $b_{QCD}^{R,D}=11-\frac{2}{3}5$ for $m<2m_t$ with the number of 
effective flavors $n_f=5$. A similar argument is also applicable to $b_{EM}^{R,D}$.

\begin{table}
\begin{tabular}{l|l|lcl}
\hline
  $R$ & $D,R$ & $D,R,H$  \\
\hline
 $\frac{1}{kL}$ & $+\frac{\alpha_s}{2\pi}(11-\frac{2}{3}6)$  & $+ \frac{\alpha_{s}}{2\pi}F_t$ ;
 & $F_t =$ & $\left\{ \begin{array}{cr} \frac{2}{3} & \ \ \ \ \ \ m<2m_t\\ 0 & 2m_t<m \end{array} \right.$\\    
\hline
 $\frac{1}{kL}$ & $+ \frac{\alpha}{2\pi}\left(\frac{19}{6}-\frac{41}{6}\right)$
 & $+\frac{\alpha}{2\pi}\left( \frac{8}{3}F_t-F_W \right)$;
 & $\frac{8}{3}F_t-F_W=$ & $\left\{ 
 \begin{array}{cr} -\frac{47}{9} & m<2m_W \\ 
 -\frac{2}{9} & \ \ \ 2m_W<m<2m_t\\ -2 & 2m_t<m 
 \end{array}\right.$    \\
\end{tabular}
\caption{$gg$ and $\gamma\gamma$ couplings, $\frac{1}{kL}+\frac{\alpha_{s}}{2\pi}b_{QCD}^i$(2nd row) 
and $\frac{1}{kL}+\frac{\alpha}{2\pi}b_{EM}^i$(3rd row), 
 of $DHR$ scalars:
Only the third column contributes for $H$ where
$F_t(F_W)$ represent the triangle-loop contributions of top quark and $W$ boson
which are given in ref.\cite{Hunter,Anatomy,book}.
For $D$, both second and third columns contribute where the second column represents the
trace-anomaly.
For $R$ the first column ($1/kL$) also contributes. 
It comes from the bulk field coupling,
The volume of the 5th dimension is taken to be $kL=35$ in $R$, 
while we can represent $D,H$ with $(1/kL)\rightarrow 0$. 
}
\label{tab1}
\end{table}

\begin{figure}[htb]
\begin{center}
\resizebox{0.7\textwidth}{!}{
  \includegraphics{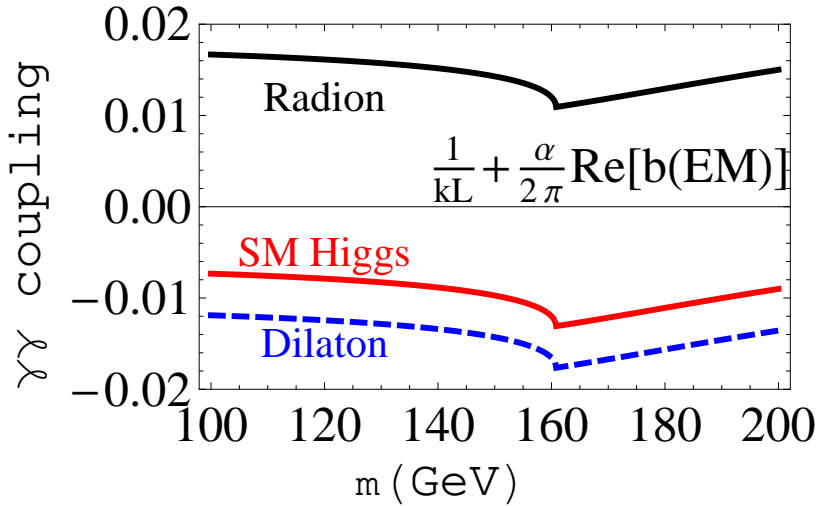}
}
%% If not, use
%%\vspace{5cm}
\end{center}
\caption{The real part of the $\gamma\gamma$ couplings,
$\frac{1}{kL}+\frac{\alpha}{2\pi}b_{EM}^i$ for $i=H,D,$ and $R$.
The bulk coupling term is given by $kL=35$ for $R$, while $(1/kL)\rightarrow 0$ for $D$ and $H$.
The $D$ has an additional contribution $-\frac{11}{3}$ in $b_{EM}^D$ compared to $b_{EM}^{H}$.
The $R$ has a contribution from the bulk field coupling which destructively interfere with 
the term of $b_{EM}^R =b_{EM}^D$.
The imaginary parts contribute above the $m>2m_W$ and they only give subleading contributions.
}
\label{fig1}
\end{figure}

The real part of the $\gamma\gamma$ couplings are given in Fig.~\ref{fig1}.
The destructive interference between the bulk coupling term $1/kL$ and the $b_{EM}^R$ term 
is due to the opposite sign, 
and this yields the very different shape of $\sigma(\gamma\gamma)/\sigma(WW)$
versus $m$:
The cusp at $m = 2m_W$, which comes from the $WW$ threshold effect,
constructively contributes for $D$ and $H$, and destructively for $R$. 
This behavior is seen in Fig.~\ref{fig2}.

$L_{AV}$ describes the $Z\gamma$ decays. The effective couplings $b_{Z\gamma}^i$ are given by
\begin{eqnarray}
b_{Z\gamma}^i &=& -A_W-A_F+\frac{b_{Z\gamma}}{{\rm sin}\theta_W {\rm cos}\theta_W},\ \ \ 
b_{Z\gamma} = \frac{19}{6} + \frac{11}{3}{\rm sin}^2\theta_W
\label{eq12}
\end{eqnarray}
where $R,H,D$ have both $A_W$ and $A_F$ terms from the triangle-loop contributions of $W$ and SM fermions,
respectively. Their explicit forms are given in refs.\cite{Hunter,Anatomy}. 
$A_F$ is negligible compared to $A_W$. 
The third term comes from the trace anomaly of $T_\mu^\mu(SM)$ and it contributes to $R,D$, but not to $H$. 
We can check that particles with heavier thresholds than $m_D$ or $m_R$ decouple also in $Z\gamma$.
To a good approximation the bulk-field coupling of $R$ gives 
no contribution to $Z\gamma$\cite{Ren}.

\begin{figure}[htb]
\begin{center}
\resizebox{0.53\textwidth}{!}{
  \includegraphics{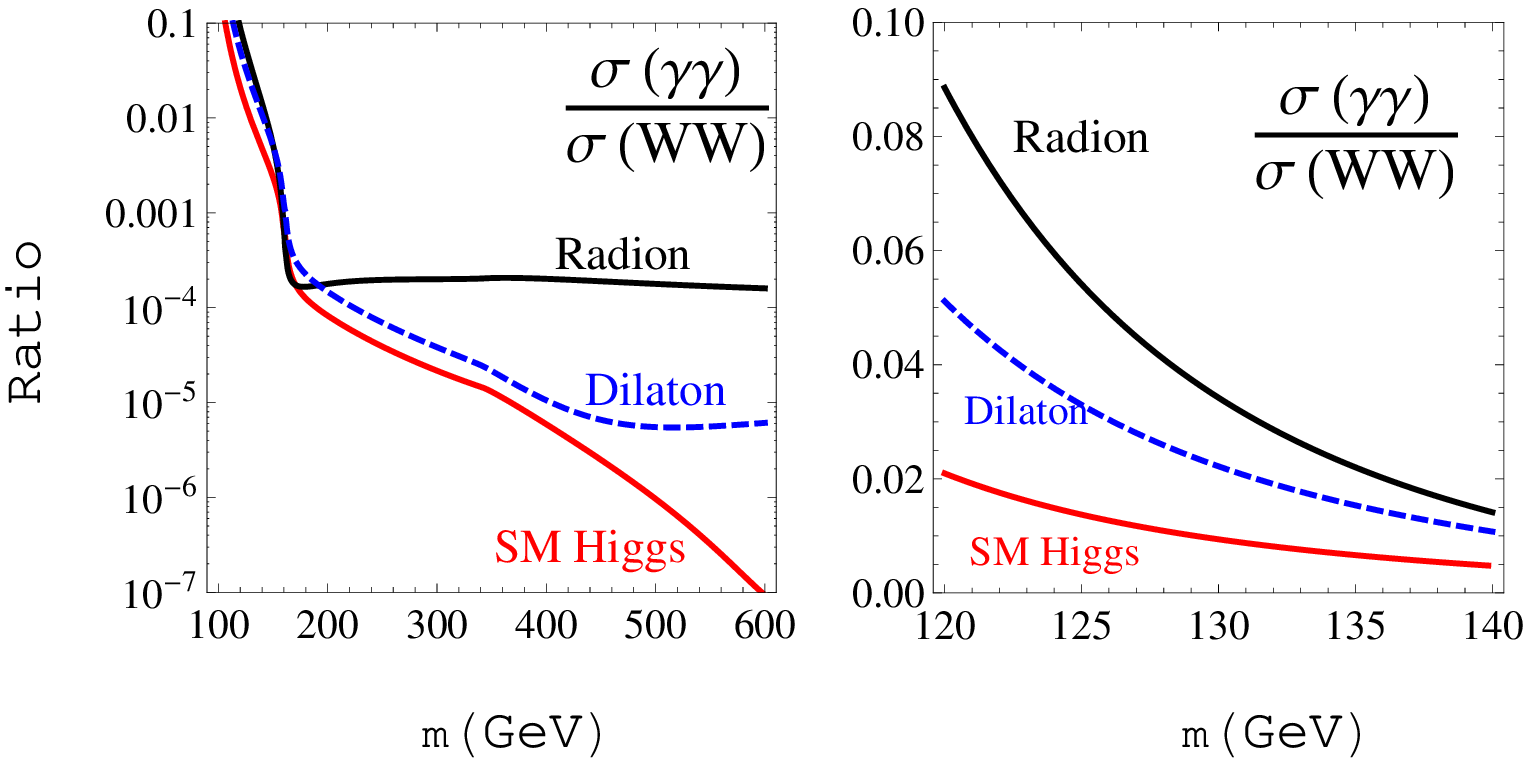}
}
\resizebox{0.53\textwidth}{!}{
  \includegraphics{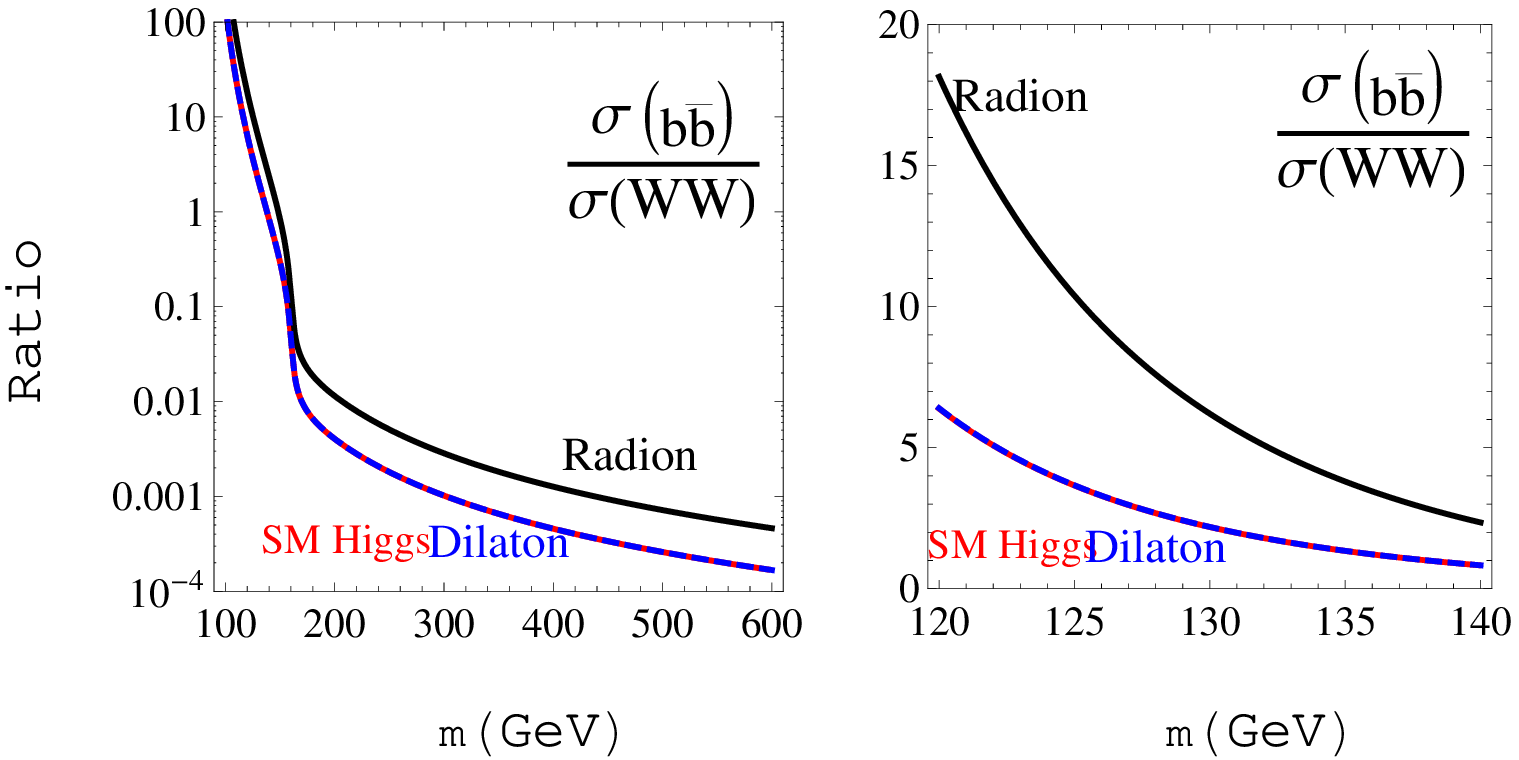}
}
\resizebox{0.53\textwidth}{!}{
  \includegraphics{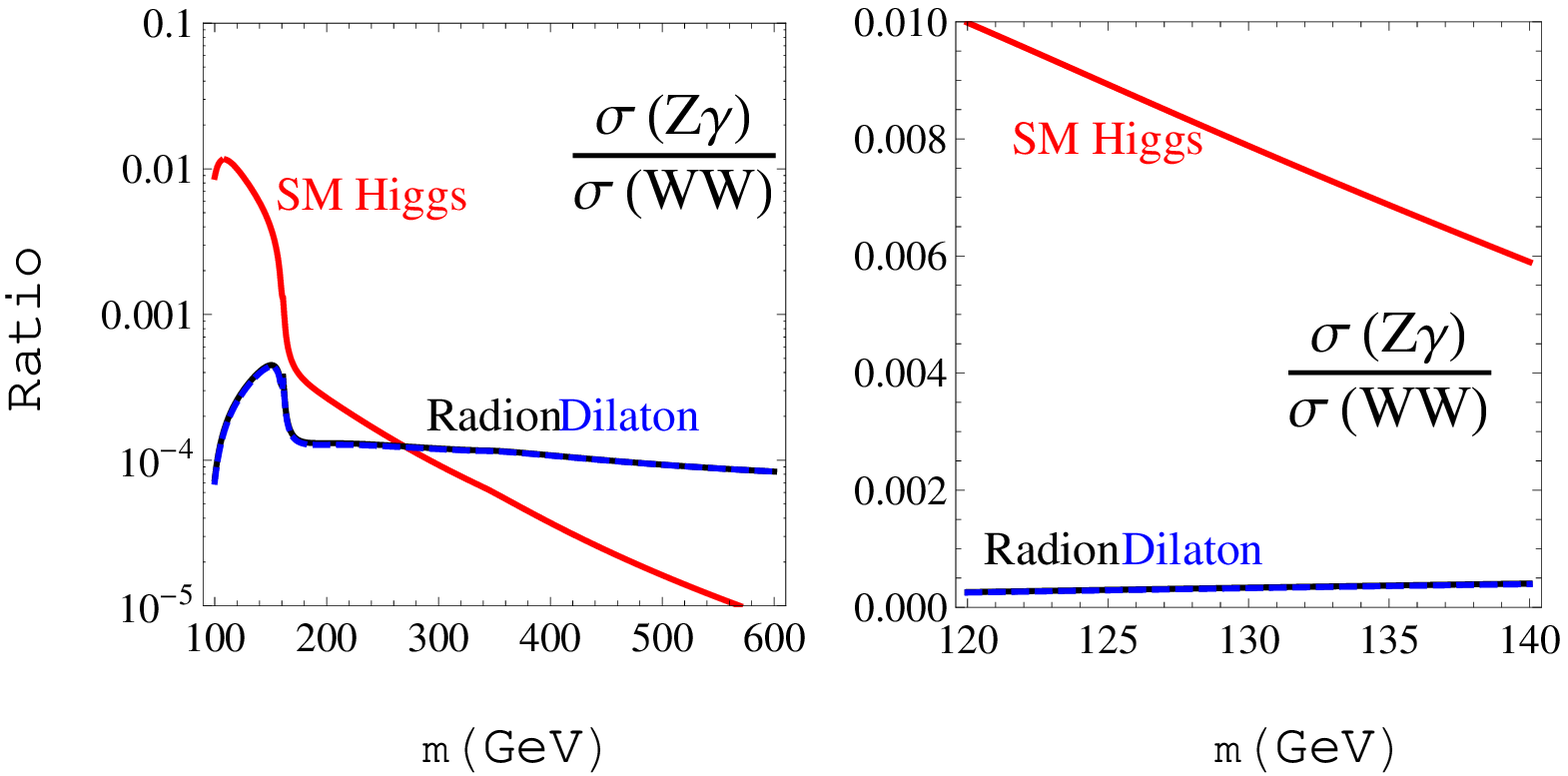}
}
%% If not, use
%%\vspace{5cm}
\end{center}
\caption{The cross section ratios, $\sigma(\gamma\gamma)/\sigma(WW)$ and
$\sigma(b\bar b)/\sigma(WW)$ versus the mass of the scalar: $H,R,D$.
For the radion(black solid), the dilaton(blue dashed), and 
the SM Higgs (red dotted) for $\sigma(\gamma\gamma,b\bar b,Z\gamma)/\sigma(WW)$(upper, middle, lower figures).
The ratios are independent of the values of the model parameters, $F_R$ and $F_D$.
$H$ and $D$ have the same value of $\sigma(b\bar b)/\sigma(WW)$ 
while for $R$ it can be different, since the ratio is proportional to the square of the parameter 
$I_b^R$ that is taken to be  1.66 as an example. $\sigma(Z\gamma)/\sigma(WW)$ of $R$ and $D$
are different from that of $H$ due to the trace-anomaly contribution.
}
\label{fig2}
\end{figure}

\noindent\underline{\it $\sigma(\gamma\gamma)/\sigma(WW)$ ratio}\ \ \ 
From $L_{\rm eff}$ in Eq.~(\ref{eq2}), we can calculate the partial widths $\Gamma$ of $H,\ R,$ and $D$.
They are proportional to the inverse squares of the overall constants $F_i$, but the values of 
$F_R$ and $F_D$ are presently unknown. 
However, the $\sigma(\gamma\gamma)/\sigma(WW)$ ratios\cite{book} 
are independent of these VEVs.
Figure~\ref{fig2} shows the ratios $\Gamma (\gamma\gamma)/\Gamma (WW)
=\sigma (\gamma\gamma)/\sigma (WW)$ (upper figure),
the ratios $\Gamma (b\bar b)/\Gamma (WW)=\sigma (b\bar b)/\sigma (WW)$ (middle figure),
and the ratios $\Gamma (Z\gamma)/\Gamma (WW)=\sigma (Z\gamma)/\sigma (WW)$ (lower figure),
for $R$ and $D$ of the same mass. They are compared with those of $H$ of the same mass.

As can be clearly seen in Fig.~\ref{fig2},
we can differentiate the three scalars, $R,D,H$, by 
observing the ratio $\sigma (\gamma\gamma)/\sigma (WW)$ .
In Fig.~\ref{fig2} the slope changes around $m\simeq 2m_W$ since $\Gamma (WW)$ steeply decreases 
below the $WW$ threshold. 
The $R$ gives an almost constant ratio in $m_R > 2m_W$ becuse of the contribution from the 
bulk coupling term $1/kL$ which is energy-independent. 
The drastic change in slope of the ratio of $R$ near $m\simeq 2m_W$ occurs from the interference 
between this bulk coupling and the trace anomaly term. See, Fig.~\ref{fig1}.

The ratio $\sigma (b\bar b)/\sigma (WW)$ of $R$ can differ from $H$ and $D$, 
because of the parameter $I_b^R$,
so measuring this quantity is also helpful to distinguish $R$ from the other two scalars.  

The ratio $\sigma (Z\gamma)/\sigma (WW)$ of $R$ and $D$ can differ from $H$, 
because of the trace anomaly contributions. 
This channel is helpful to determine the coupling form of the signal. 
%Despite the small fraction of $Z$ decays to lepton 
%pairs ($6.7\%$ for $e^-e^++\mu^-\mu^+$), 
It may be possible to detect it
by focusing on the monochromatic photon spectrum from $H\rightarrow Z\gamma$.

\noindent\underline{\it Total widths and decay branching fractions (BF)}

\begin{figure}[htb]
\begin{center}
\resizebox{0.7\textwidth}{!}{
  \includegraphics{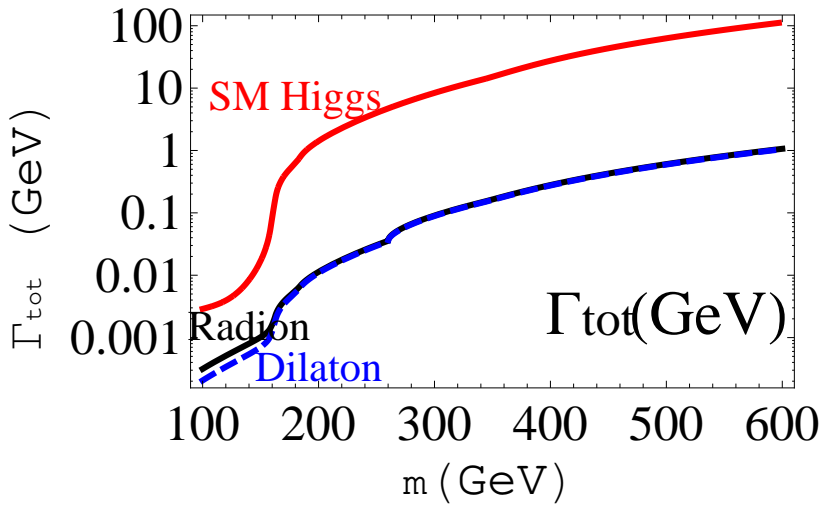}
}
%% If not, use
%%\vspace{5cm}
\end{center}
\caption{The total widths (GeV) of the radion $R$(black solid), dilaton $D$(blue dashed), and SM Higgs $H$(red dotted).  
$\Gamma_{\rm tot}^R,\Gamma_{\rm tot}^D$ scale with $(v/F_R)^2,(v/F_D)^2$, respectively, where
$F_R$ and $F_D$ are commonly taken to be 3~TeV.  
}
\label{fig3}
\end{figure}

The total widths of $R,D,H$ are given in Fig.~\ref{fig3}.
$\Gamma_{\rm tot}^R(\Gamma_{\rm tot}^D)$ scale with $(v/F_R)^2\ ((v/F_D)^2)$ where
$F_R=F_D=3$~TeV are taken, and the $R$ and $D$ widths are about two orders of magnitudes smaller than the $\Gamma_{\rm tot}^H$
with the same mass.  

\begin{figure}[htb]
\begin{center}
\resizebox{0.68\textwidth}{!}{
  \includegraphics{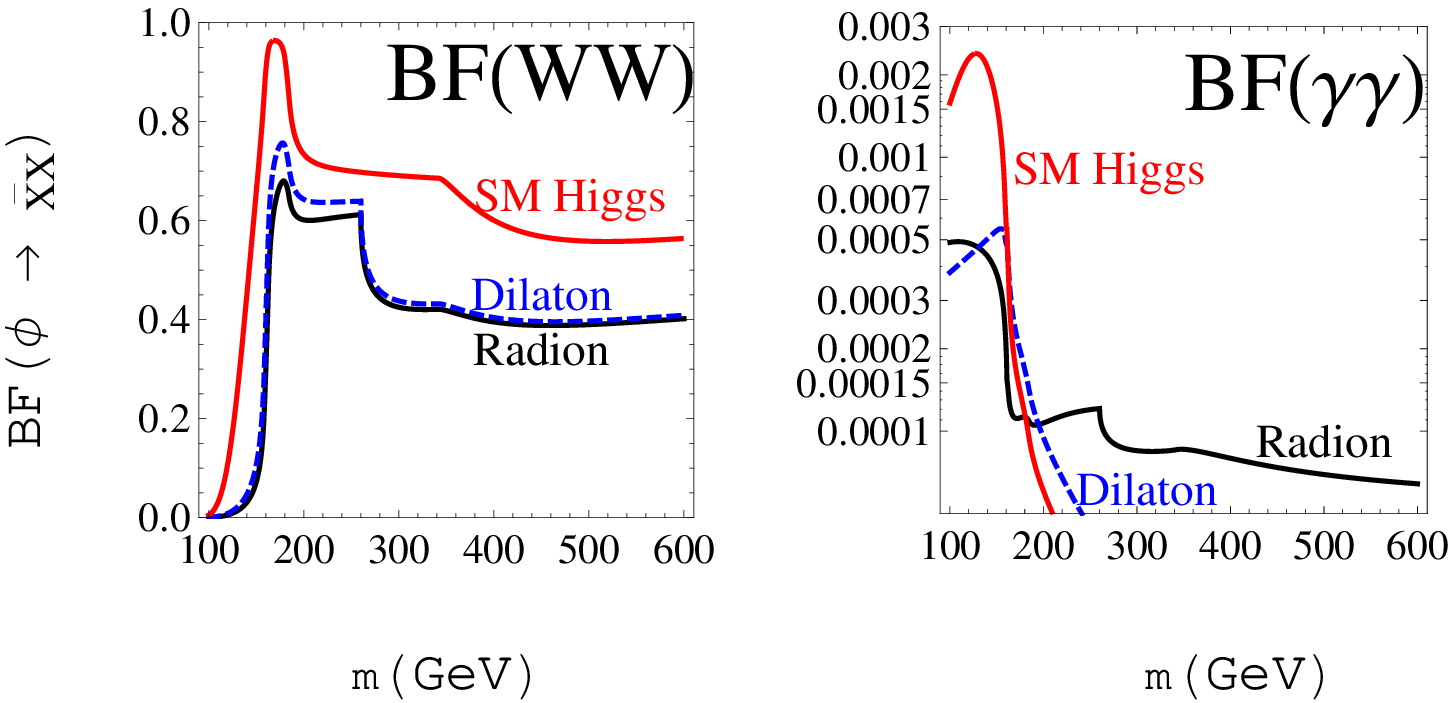}
}
\resizebox{1.0\textwidth}{!}{
  \includegraphics{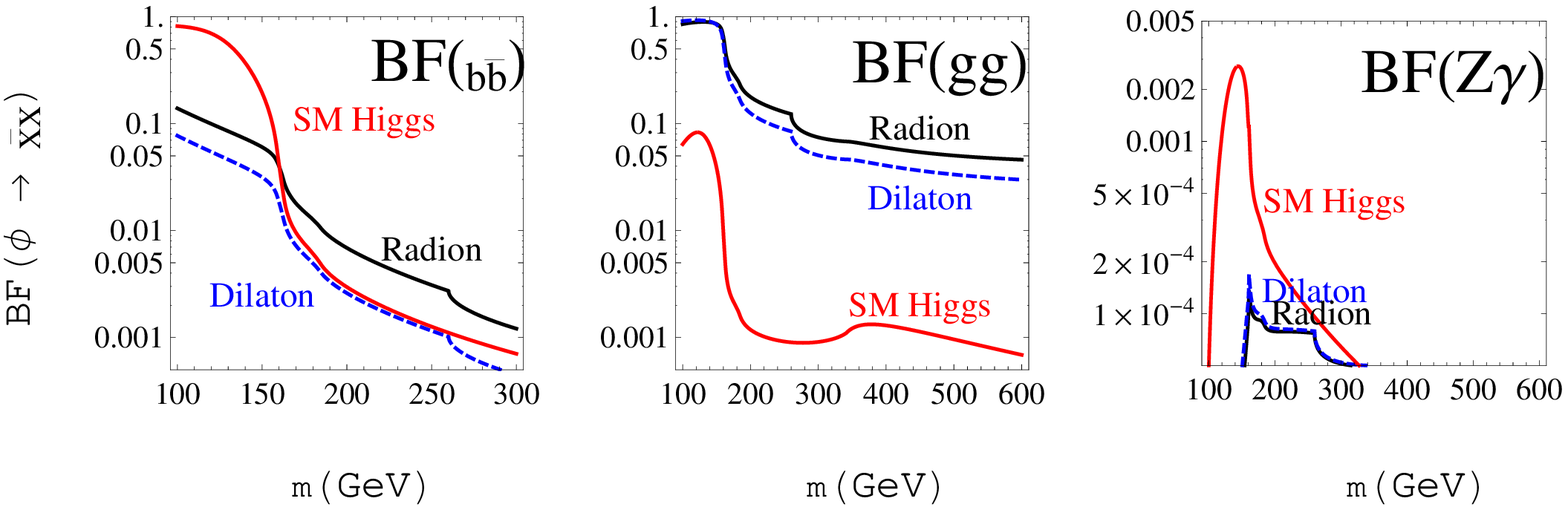}
}
\end{center}
\caption{Branching fractions $(BF)$ of the decays to $\bar XX=WW,\gamma\gamma,\bar bb,gg,Z\gamma$:
For the radion $R$ (black solid), the dilaton $D$ (blue dashed), and 
the SM Higgs $H$ (red dotted) in each figure.
This result is independent of the model parameters,
$F_R$ for the radion and the $F_D$ for the dilaton.
For $BF(\bar bb)$,
$I_b^R$ is taken to be 1.66 as an example. 
}
\label{fig4}
\end{figure}

The branching fractions ($BF$) of the decays to $\bar XX=WW,\gamma\gamma,\bar bb,gg,Z\gamma$ are compared 
in Fig.~\ref{fig4}, where the $K$-factor in NNLO\cite{Kfact} is considered for $gg$.
 BF($\gamma\gamma$) shows very delicate structures.
$BF(H\rightarrow \gamma\gamma)$ is the largest at $m<2m_W$, since the $H$ has the smallest couplings to
$gg$ and the main $H$ decay mode in this energy region is $b\bar b$. 

\noindent\underline{\it Concluding Remarks}\ \ \ \ 
The measurement of the ratio\cite{ratio} $\sigma(\gamma\gamma)/\sigma (WW)$ provides a decisive way to
differentiate the radion $R$, the dilaton $D$, and the SM Higgs $H$.
It is only necessary to count the event numbers of $\gamma\gamma$ and $WW$ decays of an observed signal.
This method is independent of the values of the model-parameters,
the VEVs $F_{R}$ and $F_{D}$.
It applies to both the LHC and Tevatron experimental searches.

The scalars are also expected to be produced in $W/Z$ associated production,
$W^*\rightarrow W\varphi^i$ and $Z^*\rightarrow Z\varphi^i$. The production cross section 
$\sigma_{\rm assoc.}(D)$ and  $\sigma_{\rm assoc.}(R)$ are smaller than $\sigma_{\rm assoc.}(H)$,
respectively, by the factors
$(\frac{1}{F_D})^2$ and $(\frac{1}{F_R})^2$,
which are $\sim 0.01$ in the $F_D\sim F_R\sim 3$~TeV case. 
This small cross section of associated production also can be used to 
differentiate the $R$ and $D$ from $H$\cite{Logan}.  

The production of $D$ and $R$ via the $WW,ZZ$ fusion subprocess is much smaller than that of $H$,
due to their relatively smaller decay widths to $WW$ and $ZZ$.

We may also consider the scenario that both $D$ and $H$ (or $R$ and $H$) exist with comparable masses 
in the region $m\sim 125$~GeV, where
the on-going Higgs search data show some excess over the expected SM cross section.
At this mass both $D(R)$ and $H$ have very narrow widths, and their resonance peaks will be smeared
by experimental resolution 
into one with twice the production cross section, even with the mixing of scalars taken into account.
In this case the $\sigma(\gamma\gamma)/\sigma(WW)$ ratio will be intermediate between the single-state values. 
Another possible scenario is that $R$ or $D$ mix\cite{GRW,mix1,mix2,mix3} 
with $H$. Then, the lighter scalar can have a mass below 100~GeV
and its production will be suppressed compared to that of the SM Higgs. 

For dilaton or radion masses much larger than $2m_W$, the narrow width makes discovery in $WW$ and $ZZ$ 
easier than for the SM Higgs\cite{Logan}. 

Finally,
our study applies also to generic singlet models\cite{NMSSM}.
The singlet decouples from SM particles 
and the phenomenology is dependent on the amount of mixing of $H$ with the singlet scalar. 
The $H$ production cross section can be significantly smaller by the mixing effect, and thus, 
a low-mass Higgs boson with $m_H<100$~GeV also become possible.

\noindent\underline{\it Acknowledgements}

We thank  Professors Bill Bardeen and Prof. Misha Stephanov for discussions.
M.I. is very grateful to the members of phenomenology institute of University of Wisconsin-Madison for hospitalities.
This work was supported in part by the U.S. Department of Energy under grants No. DE-FG02-95ER40896 and
DE-FG02-84ER40173, 
in part by KAKENHI(2274015, Grant-in-Aid for Young Scientists(B)) and in part by grant
as Special Researcher of Meisei University.

% The \nocite command causes all entries in a bibliography to be printed out
% whether or not they are actually referenced in the text. This is appropriate
% for the sample file to show the different styles of references, but authors
% most likely will not want to use it.
\nocite{*}

\bibliography{apssamp}% Produces the bibliography via BibTeX.

\end{document}